# Navigating Governance Paradigms: A Cross-Regional Comparative Study of Generative AI Governance Processes & Principles

**Jose Luna, Ivan Tan, Xiaofei Xie, Lingxiao Jiang**

Singapore Management University

joseluis.lc.2023@phdcs.smu.edu.sg, ivantan@smu.edu.sg, xfxie@smu.edu.sg,lxjiang@smu.edu.sg

## Abstract

As Generative Artificial Intelligence (GenAI) technologies evolve at an unprecedented rate, global governance approaches struggle to keep pace with the technology, highlighting a critical issue in the governance adaptation of significant challenges. Depicting the nuances of nascent and diverse governance approaches based on risks, rules, outcomes, principles, or a mix, across different regions around the globe, is fundamental to discern discrepancies and convergences, and to shed light on specific limitations that need to be addressed, thereby facilitating the safe and trustworthy adoption of GenAI. In response to the need and the evolving nature of GenAI, this paper seeks to provide a collective view of different governance approaches around the world. Our research introduces a Harmonized GenAI Framework, "H-GenAIGF", based on the current governance approaches of six regions: (European Union (EU), United States (US), China (CN), Canada (CA), United Kingdom (UK), and Singapore (SG)). We have identified four constituents, fifteen processes, twenty-five sub-processes, and nine principles that aid the governance of GenAI, thus providing a comprehensive perspective on the current state of GenAI governance. In addition, we present a comparative analysis to facilitate identification of common ground and distinctions based on coverage of the processes by each region. The results show that risk-based approaches allow for better coverage of the processes, followed by mixed approaches. Other approaches lag behind, covering less than 50% of the processes. Most prominently, the analysis demonstrates that amongst the regions, only one process aligns across all approaches, highlighting the lack of consistent and executable provisions. Moreover, our case study on ChatGPT reveals process coverage deficiency, showing that harmonization of approaches is necessary to find alignment for GenAI governance.

## Introduction

In the rapidly evolving landscape of Artificial Intelligence (AI), the advent of GenAI marks a significant shift, heralding a new era of technological sophistication and societal transformation. With such power and fast evolution, governments are trying to adapt their AI governance approaches to incorporate the unique challenges that GenAI poses (Samuelson 2023; Ferrara 2024; Bird, Ungless, and Kasirzadeh 2023). GenAI, characterized for its ability to generate new and adaptive outputs, stands apart from previous AI technologies. These outputs could be in the form of text, images, audio, amongst others. The pioneering work on Generative Adversarial Networks (GANs) (Goodfellow et al. 2014) marked the beginning of the transformative journey of GenAI. Thereafter, the subsequent development of Transformer models (Vaswani et al. 2017) further expanded GenAI's capabilities, enabling applications that range from content creation to decision-making processes, deeply influencing industries. One of such applications is ChatGPT, introduced in November 2022, reaching 100 million users in record time (Wu et al. 2023), underscoring the profound societal influence of GenAI. The impact extends beyond technological innovation, seeping into the fabric of everyday life, altering the landscape of work (Gmyrek, Berg, and Bescond 2023), creativity (Doshi and Hauser 2023; Figoli, Mattioli, and Rampino 2022), and education (Lim et al. 2023; Lagran, Searson, and Trumble 2024), thereby demonstrating the necessity for robust governance frameworks. However, AI governance is struggling to adapt at the same pace. Although, adaptations to AI governance are being made at the time this work was produced (IAPP 2024). To set the premise, it is important to understand current developments of GenAI governance within the regions studied.

To begin with, the EU's AI Act represents a pioneering effort to integrate GenAI into a harmonized regulatory approach (risk-based) for EU member states (EU 2024a). Their approach is geared towards ensuring that AI systems are human-centric, trustworthy and attempts to manage the risks associated with GenAI by providing transparency obligations, making sure that providers of such models take into account safeguards against the risks associated with said models and their integration into other systems ( European Parliament 2024).

With a similar risk-based approach, the US introduces an executive order putting an emphasis on safety, security, and trustworthiness in AI (The White House 2023). Although there is a minimal focus on GenAI, there are numerous mentions of complementary documents and when they will be released. One such document, the draft risk management framework for GenAI (NIST 2024), provides a use-case profile that guides organizations in managing AI risks, incorporating trustworthiness into AI design and adapting to cross-sectoral applications. Moreover, US has introduced different bills that target GenAI, tackling copyright disclosure (US Congress 2024a), preventing misuses of deceptive content (US Congress 2023b,a), and protecting individuals' rights over their digital likenesses (US Congress 2024b).

In comparison, the UK's approach (outcome-based) takes a proactive and innovation-driven perspective (DSIT 2023). GenAI governance is emphasized, particularly within government and public sector applications, ensuring meaningful human control (UKGov 2024).

CA's principle-based approach undertakes initiatives to ensure that GenAI systems are safe, respectful of human rights, and trustworthy (Government of Canada 2022). For GenAI, CA introduced a code of conduct (Government of Canada 2023) which is designed to guide organizations in implementing risk management practices.

In Asia, particularly in CN, the approach to GenAI governance is characterized by a structured set of rules designed to harmonize innovation with comprehensive oversight. This approach (rule-based) prioritizes alignment with national values, ensuring that technological advancements do not undermine national security or societal stability (Cyberspace Administration of China 2023). In SG, the proposed governance framework for GenAI (IMDA and AIVF 2024) stands out for its well-structured and practical approach (risk-principle based) for responsible implementation of GenAI, ensuring that its development and applications are beneficial, safe, and aligned with the country's broader societal and economic objectives (Smart Nation 2023).

Although various regions have developed distinct regulatory guidelines for GenAI, which specify the GenAI-related actions or processes being "regulated", there remains a limited exploration of what processes are covered by these guidelines, how they are regulated or evaluated, the differences across regions, and how these provisions can be applied to GenAI systems. To address these gaps, this paper presents an empirical study on the governance approaches from six regions and introduces a novel Harmonized GenAI Governance Framework (H-GenAIGF). This framework includes: 1) A taxonomy of GenAI governance processes across common constituents; 2) Governance principles; 3) A mapping between the principles and the processes/sub-processes; and 4) A cross-regional comparative analysis on GenAI processes coverage.

In summary, this paper makes the following contributions:

- We conduct an empirical analysis of GenAI regulatory guidelines from six regions and developing a Harmonized GenAI Governance Framework.
- We perform a comparative analysis on the coverage of GenAI processes from different regions.
- We apply the framework in a case study on the ChatGPT system to illustrate its practical implications.

This study helps mitigate governance gaps, facilitating the alignment with technological advancements, promoting ethical practices, and safeguarding societal values. The H-GenAIGF not only aids in identifying governance gaps and inconsistencies but also provides a harmonized blueprint for policymakers, industry leaders, and other stakeholders.

## Related Work

Although GenAI is relatively new in the AI scene, there is a vast amount of research focusing on how to apply the technology, from large language models (LLMs) (Hadi et al. 2023; Myers et al. 2024) to large video models (Selva et al. 2023). However, when it comes to GenAI governance, more effort is needed to establish synergy between the technical and governance aspects of GenAI. This section bifurcates into *Academic Research for Governance* and *Governmental Efforts*. Academic research and government efforts play crucial roles in the governance of GenAI, and they synergistically inform and shape responsible and trustworthy AI deployments.

### Academic Research for Governance

Back in 2017, AI governance was an important agenda and knowledge gap identified by many (Dafoe 2018; Birkstedt et al. 2023). Despite the tremendous opportunities AI offers, the potential consequences of its risks, both severe and urgent, have prompted calls from government and business leaders for policy guidance. However, academic engagement with the AI revolution was relatively limited.

At that time, research on AI governance was critically needed to establish global norms, policies, and institutions for the beneficial development and use of advanced AI. By 2017 and 2019 respectively, ISO and IEEE emerged as leading bodies for AI standards (ISO 2017; IEEE 2019), focusing mainly on market efficiency and ethical concerns. However, these standards risked overlooking broader policy objectives such as promoting responsible deployment and ensuring safety in fundamental research. Notably, prominent AI research organizations concerned with these policy goals were largely absent from these standardization efforts.

Thus, recommendations (Cihon 2019) included building institutional capacities, such as "Leading AI labs should build institutional capacity to understand and engage in standardization processes" and "Standards should be used as a tool to spread a culture of safety and responsibility among AI developers". The team in (Pagallo et al. 2019) also suggested 14 priority actions, a "SMART" model of governance, and a regulatory toolbox to advance the research in AI governance research. In 2022, researchers (Gordon, Rieder, and Sileno 2022) also did work on mapping values in AI governance. However, the standards and values suggested by these papers were not directly related to GenAI governance processes specifically, but rather the general field of AI. Nevertheless, each of the approaches done in all of these works have all played a huge role in the advancement of AI governance, to the point it is at today. There is much to learn from these contributions, especially when trying to govern an emerging field like GenAI. Other academic studies have also created taxonomies for the given context (Arya et al. 2019; Samoili et al. 2020; Dellermann et al. 2021; Wang, She, and Ward 2021). Such taxonomies facilitate governance by structuring information and concepts, promoting clarity, consistency, and standardization across entities.

Furthermore, many advanced techniques have been proposed to evaluate the quality, reliability, and security of models (Sun et al. 2024; Wang et al. 2023a; Du et al. 2019; Xie et al. 2019a, 2022, 2021, 2019b,a). These techniques can serve as valuable tools for AI governance. However, there remains a gap between governance and these technical tools.

**Governmental Efforts**

Apart from the Governmental Documents (GD) mentioned in the introduction, there have been a range of developments for GenAI governance. For instance, initiatives have emerged to understand the opportunities and risks of LLMs and multimodal foundation models (MFMs) (Bell et al. 2023). Other works foster an AI-friendly ecosystem, where guidance for innovation is presented in the form of practical use cases of GenAI (United Arab Emirates 2023).

In specific sectors like education and research, UNESCO has issued guidelines for GenAI (UNESCO 2023a), discussing eight major controversies surrounding its use in education. The guidelines also address concerns related to access and equity, human connection, intellectual property, psychological impact, and hidden bias and discrimination. Additionally, they provide a framework for employing GenAI in education and research. While these guidelines represent progress, they fall short in offering actionable processes due to a lack of specificity in the recommendations.

Regarding research, the recent EU's initiative on responsible use of GenAI for research (EU 2024b) emphasizes four principles underpinning the recommendation. Reliability, honesty, respect, and accountability guide the responsible use of GenAI in research, ensuring the quality, transparency, and social impact of scientific activities, covering three main stakeholders: researcher, research organizations, and funding organizations. However, no emphasis on other crucial principles such as auditability is mentioned. It can be said that the EU guidelines for use of GenAI in research are a complement to prior work such as the AI's Ethics Guidelines for Trustworthy AI (EU 2019), the EU AI Act (EU 2024a), and in a broader scale align with the Recommendation on the Ethics of AI (UNESCO 2023b).

As a combined effort, the Group of Seven's (G7) tackled advanced AI systems (G7 2023), including GenAI systems, to provide a set of "guiding principles" that guide organizations in developing and using advanced AI systems, emphasizing the need for safety, security, and trustworthiness. In total, 11 recommendations have been set, such as employing diverse internal and external testing measures, fostering responsible information sharing, and prioritizing transparency and research to mitigate societal risks, among others. Thus, organizations are urged to implement robust governance policies and explore reliable content authentication like watermarking. Moreover, the "guiding principles" align with the Organization for Economic Co-operation and Development (OECD) AI Principles (OECD 2024), and emphasizes the need of international collaboration.

As shown in the related work, there exists a conspicuous absence of unified approaches for GenAI governance processes. This underscores the importance of our study, which endeavors to establish a harmonized framework to facilitate effective comparisons and ensure better adherence to GenAI governance processes. Such a framework is essential for identifying best practices and promoting consistency in GenAI governance.

**Methodology**

Figure 1 illustrates the methodology of our study. Our methodology consists of gathering GD from regions with differing governance approaches related to GenAI, identifying and categorizing the constituents, processes & sub-processes, the principles mapping & cross-regional analysis, and applying the framework to a real-world GenAI application.

In this paper we refer to GD as regulations, bills, acts, frameworks, and guidelines. Understanding there is a clear hierarchical order of GD, with some being more authoritative (e.g. regulations) than others (e.g. guidelines). The selection criteria for the different governance approaches was inspired from the World Economic Forum (WEF) briefing paper series (WEF 2024), where four approaches to AI governance were presented, namely, Risk, Rules, Principles, and Outcomes approaches. As the approaches stated are not mutually exclusive, we adapted these to include regions with mixed approaches. Thus, EU and US follow a risk-based approach, CN a rule-based approach, CA a principle-based approach, UK an outcome-based approach, and SG a risk-principle based approach. Furthermore, these regions also have significant impact in their specific geographical locations. As example, SG and its influence in ASEAN, given that SG has placed second in the "Government AI Readiness Index" (Hankins et al. 2023). Our study investigates how different approaches can be compared against each other, in terms of coverage of governance processes.

Furthermore, to extract and organize the key provisions or recommendations from the GD in a structured manner, we used four constituents: data, model, content generation, and ethics. These are found to be common ground for the different approaches, as they are mentioned in all regions (EU 2024a; Cyberspace Administration of China 2023; NIST 2024; Government of Canada 2023; IMDA and AIVF 2024; UKGov 2024). Additionally, they represent key aspects of GenAI by exhibiting what goes into GenAI systems (data), the inner functionality (model), the outputs (content generated), and the fundamental alignments (ethics).

Afterwards, we identified and categorized processes and sub-processes under each constituent following a manual three-cycle consideration (to be explained later). We refer to processes to the primary categories of activities within GenAI governance. Each process covers an area of governance focus that encompasses various actions, and these actions are referred to as sub-processes. Besides, sub-processes provide a detailed view of how to address adherence to a process.

Additionally, we refer to principles as essential fundamentals that drive governance adherence. Consequently, the mapping of these principles to the corresponding process/sub-process was performed. Such principles are also extracted from the GD and organized following the same three-cycle consideration. Hence, a harmonized taxonomy of governance processes for GenAI, with the corresponding governing principles is constructed. The manual analy-sis section provides further details on how the key provisions/recommendations of the GD were organized.

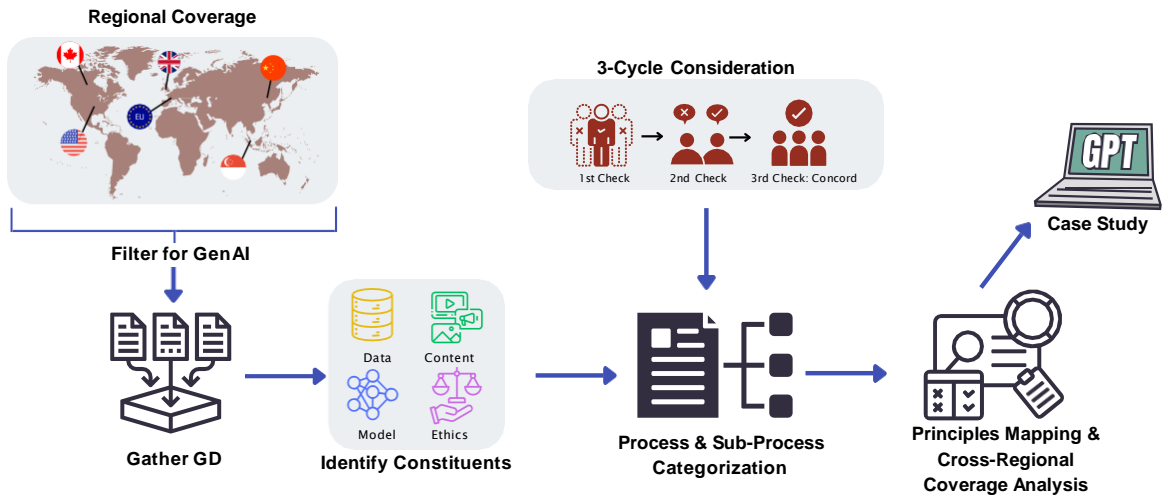

Figure 1: Methodology Overview

Based on our harmonized taxonomy, we perform a comparative analysis of the processes/sub-processes across different regions, in particular, their coverage by the GD of different regions, illustrating the commonalities, differences, and governance gaps across regions. Lastly, to demonstrate the severity and known impact of GenAI, not only to policymakers but also to society, we conducted a case study on ChatGPT-3.5&4 covering the different constituents, while focusing on content generation as it is the constituent that shows most significant impact to society (Partadiredja, Serrano, and Ljubenkov 2020).

### Document Collection

We collected regulatory approaches such as acts and bills with the intention of identifying regions that are enforcing, or planning to enforce, certain provisions to GenAI. As mentioned earlier, all these documents are grouped as GD, and were obtained directly from reputable sources such as the official web sites of governmental agencies (US Congress, White House, EU Parliament, Government of Canada, Cyber Space Administration of China, UK Government, Infocomm Media Development Authority of Singapore) which have been previously referenced. The GD collected are reasonably representative of globally distributed regions for our cross-regional study of GenAI governance.

A filter was included in our collection process (Beale et al. 2014). We did not include GD that only cover traditional AI, unless GenAI is an expansion or complement to said GD. As an example, in the case of EU AI Act, the latest update introduced the provisions to regulate general purpose AI, and these systems are also required to adhere to the EU AI Act provisions previously approved, as well as the updated provisions for general purpose AI. A total of 15 documents of varied lengths (from 2 pages to 459 pages) were collected. In addition, to explain the governance processes more in detail, we collected relevant academic papers to compliment the lack of detailed explanation from the GD.

### Manual Analysis & Case Study Selection

To construct our framework we first identified four constituents: Data, Model, Content Generation, and Ethics, as a recurrent theme in all regions' approaches. This allows us to further categorize the processes and sub-processes under corresponding constituents. To identify and categorize processes and sub-process, we undertook an extensive, manual and labour-intensive analysis of all provisions and recommendations from the GD.

As GD often contains ambiguous or poorly-defined concepts and terms, and to avoid biases, we followed a three-cycle consideration to extract each provision/recommendation and its placement under their corresponding constituent or process (in the case of the sub-processes). Initially, the first person extracts and places the relevant information under a specific constituent or process. Subsequently, another person reviews and considers the placement. Finally, if a disagreement arises during the second stage regarding the interpretation or placement of a provision/recommendation, a third person will review the matter with the objective of achieving concord. This process is repeated until all provisions/recommendations are accurately organized under their corresponding constituents/processes. Similar approach is adopted for identifying the principles of the processes.

Based on the organization of the different components of our framework, we perform the comparative analysis incorporating selected regions.

At last, to demonstrate the usefulness of our framework, we included a case study on ChatGPT 3.5&4. We selected ChatGPT due to its massive adoption, economic and societal impact (Baldassarre et al. 2023; Zarifhonarvar 2023), and regulating such models is an active research topic (Hacker, Engel, and Mauer 2023) and should be illustrated how our governance framework can be applied to them.

# H-GenAIGF

The days where enforceable practices for the development of AI systems may be closer (Mökander et al. 2022; Erdélyi and Goldsmith 2018). However, many developers, AI providers and/or end users may find navigating the convoluted nature of governance approaches burdensome. Thus, the need for a harmonized framework to understand the processes of GenAI governance becomes essential. Not only for policymakers, but also for developers and the end users. In previous section, we have presented different approaches to govern GenAI. Although some similarities are shared between them, there are clear differences in the level of regulatory control, intent, and processes coverage.

To develop a harmonized framework for GenAI processes, the examination of various governance approaches must be explored. To this end, a panoramic view of how different regions are tackling GenAI governance is needed. Therefore, we approach this by employing a targeted analysis that covers SG, CN, EU, US, UK, and CA on how well they cover the identified processes and sub-processes. In selecting these regions, we focused on jurisdictions that have different approaches to AI governance. These regions were selected as described in the methodology section.

This section will dive into the description of the identified 15 processes, 25 sub-processes, and their categorization under each of the 4 constituents. For further supporting explanation, refer to our website[1]. Figure 2 provides a more comprehensive view of the H-GenAIGF, including the mapping of the principles to the processes and sub-processes, and further commentary will be given in later sections.

## Data

Under the data constituent, we identify three main processes covered in the GD across various countries:

- *Data Acquisition* - This process is divided into two main sub-processes. First, data sourcing, encompasses collecting personal data, protecting intellectual property rights, and obtaining legal and representative datasets (Choi, Jeon, and Kim 2019; Borovicka et al. 2012). Combined, the process involves strictly adhering to applicable legal frameworks when gathering information that can directly identify individuals. In addition, for GenAI there is a heightened focus on protecting intellectual property rights (Abbott 2020), extending beyond mere copyright considerations. Furthermore, the importance of sourcing datasets that meet legal standards and are tailored to specific tasks cannot be overstated. The second sub-process, obtaining consent, is pivotal in aligning with privacy laws by clearly communicating to data subjects how their data will be utilized (Jesus and Mustare 2019; Andreotta, Kirkham, and Rizzi 2022). Although essential, only half of the studied processes cover it.
- *Data Preparation* - From the GD, there is a common focus on Annotating & Labelling to ensure that data is accurately described and categorized, and data cleaning is mentioned in the context of removing inaccuracies and process data to ensure high quality.
- *Storing & Sharing Data* - Data sharing and storing are common processes in AI development. We should ensure secure storage and the inclusion of sharing protocols that must protect data securely such as confidentiality and integrity.

## Model

There are five processes within the model constituent.

- *Model Development* - Covers selecting or creating models that originate from legal sources. Additionally, disclosing model infrastructure and architecture (Di Porto 2023; Laux, Wachter, and Mittelstadt 2024) for transparency purposes becomes essential.
- *Model Validation & Testing* - This is a critical process for GenAI governance, designed to ensure that AI models operate reliably and safely under varied circumstances. Within the process we encounter Context Testing, which assesses how models perform across different operational environments to ensure their robustness and adaptability (Hernández-Orallo 2017); Adversarial Testing, which involves challenging the model with intentionally crafted inputs that attempt to cause the model to fail, thus evaluating the resilience of AI systems against potential malicious attacks (Hannon et al. 2023). Additionally, Safety & Performance Benchmarking is conducted to measure a model's performance against established safety and efficiency metrics (Hamid et al. 2023; Raji et al. 2021). Lastly, efforts in Preventing/Identifying Adversarial Attacks are important as they focus on enhancing the security measures within AI models to detect and mitigate threats proactively (Wang, Wu, and Zheng 2024; Costa et al. 2024).
- *Model Deployment* - Pivotal process encompassing sub-processes such as the Distributing Method and Operational Integration. The Distributing Method involves the strategic release and distribution of AI models, ensuring that they are accessible in appropriate formats across various platforms and environments. Concurrently, Operational Integration refers to the seamless incorporation of AI models into existing technological infrastructures and business processes. This integration is essential for maximizing the functional capabilities of AI systems, ensuring that they interact effectively with other software and hardware components (Sohn et al. 2020).
- *Model Maintenance* - Involves regular updates and refinements to ensure AI models continue to perform optimally (Davis, Emb´ı, and Matheny 2024). This continuous upkeep helps to mitigate the risk of model degradation over time, maintaining the accuracy and relevance of AI applications.
- *Compliance & Risk Analysis* - Plays a crucial role, particularly through Context Risk Analysis and Incident Reporting. Context Risk Analysis (Wach et al. 2023) assesses the potential risks associated with deploying AI models. This analysis is fundamental in preemptively addressing potential issues that could arise from contextual dynamics. Furthermore, Incident Reporting is a critical mechanism that ensures any anomalies or failures within

---

[1]https://sites.google.com/view/h-genaif/home

| Constituent | Processes | Sub-Processes | SG | CN | EU | USA | UK | CA |
|---|---|---|---|---|---|---|---|---|
| **H-GenAIGF: Processes, Principles & Cross-regional Coverage** ||||||||||
| DATA | Data Acquisition[A] | Data Sourcing[T,F,P] | - | ✓ | ✓ | ✓ | - | - |
| | | Obtaining Consent[T,P] | X | ✓ | ✓ | ✓ | X | X |
| | Data Preparation[A] | Annotating & Labelling[T,I] | ✓ | ✓ | ✓ | ✓ | X | X |
| | | Data Cleaning[T,I] | ✓ | ✓ | ✓ | X | X | X |
| | Storing & Sharing data[S] | Store | X | X | X | X | X | X |
| | | Share[T] | X | ✓ | ✓ | ✓ | X | X |
| MODEL | Model Development[I,A,S] | Sourcing Models[T] | X | ✓ | ✓ | X | X | X |
| | | Disclosing Infrastructure & Architecture[T] | - | X | ✓ | ✓ | - | X |
| | Model Validation & Testing[T,F,A] | Context Testing | ✓ | X | X | ✓ | X | ✓ |
| | | Adversarial Testing | ✓ | X | ✓ | ✓ | - | ✓ |
| | | Safety & Performance Benchmarking | ✓ | X | X | ✓ | X | X |
| | | Preventing/Identifying Adversarial Attacks | ✓ | X | ✓ | ✓ | X | ✓ |
| | Model Deployment[A,In] | Distributing method | X | X | ✓ | ✓ | X | X |
| | | Operational Integration | X | X | ✓ | ✓ | X | X |
| | Model Maintenance[A] | | ✓ | X | ✓ | ✓ | X | ✓ |
| | Compliance & Risk Analysis[A] | Context Risk Analysis | ✓ | X | ✓ | ✓ | X | X |
| | | Incident Reporting | X | X | ✓ | ✓ | X | X |
| CONTENT GENERATION | Content Validation & Moderation[I] | Assessing & Mitigating Toxicity | ✓ | X | X | X | - | X |
| | | Verifying Content Provenance | ✓ | X | ✓ | ✓ | X | ✓ |
| | | Labeling Generated Content[T] | X | ✓ | ✓ | ✓ | ✓ | X |
| | | Managing & Mitigating of Unlawful Content[F] | X | ✓ | ✓ | ✓ | X | X |
| | | Protecting Against Unlawful Content[F] | ✓ | X | ✓ | ✓ | - | X |
| | Managing Distribution & Access control[F] | | X | X | ✓ | ✓ | X | X |
| | Disclosure[T] | Providing Content Disclaimers | X | X | ✓ | ✓ | X | ✓ |
| | | Detecting GenAI Content | ✓ | X | ✓ | ✓ | - | ✓ |
| | Feedback[F,R] | | X | ✓ | ✓ | - | - | X |
| ETHICS | Ethical Alignment & Human Rights[F] | | ✓ | ✓ | ✓ | ✓ | ✓ | ✓ |
| | Ethical Design & Deployment | Ensuring Accountability & Responsibility throughout the Lifecycle[Ac] | ✓ | - | ✓ | - | X | - |
| | | Maintaining Sustainability & Environmental Responsibility[S] | ✓ | X | ✓ | ✓ | ✓ | X |
| | Upholding User Rights & Control[P] | | X | X | ✓ | X | X | X |

**Principles**:
- **T**= Transparency
- **Ac**= Accountability
- **S**= Sustainability
- **F**= Fairness
- **A**= Auditability
- **P**= Privacy
- **I**= Integrity
- **In**= Interoperability
- **R**= Responsiveness

Figure 2: Overview of H-GenAIGF

AI operations are promptly documented and analyzed (McGregor 2021). This not only aids in immediate rectification but also contributes to the iterative improvement of AI systems.

### Content Generation

This constituent is arguably the most impactful for society, as the outputs generated by GenAI models are often at the fingertips of end-users. Within this constituent, we identify four processes:

- *Content Validation & Moderation* - Process that is critical for ensuring the ethical and legal integrity of GenAI-generated outputs. This process includes vital sub-processes such as Assessing & Mitigating Toxicity (Faal, Schmitt, and Yu 2023) that rigorously tests to identify and diminish harmful content (Hartvigsen et al. 2022), adhering to governance practices and ethical standards. Next, Verifying Content Provenance maintains transparency by ensuring content authenticity (Balan et al. 2023; Hamed, Zachara-Szymanska, and Wu 2024). Afterwards, Labeling Generated Content marks outputs as machine-generated, providing crucial context for end-users (Wittenberg et al. 2024). Following this, Managing & Mitigating of Unlawful Content involves proactive strategies to prevent the production and dissemination of illegal material (Schramowski et al. 2023; Romero Moreno 2024a), while Protecting Against Unlawful Content (Gupta et al. 2023; Kilovaty 2025) rapidly addresses and rectifies any infractions, ensuring regulatory compliance and user protection.

- *Managing Distribution & Access Control* - Ensures that generated content is disseminated responsibly and equitably (Dinzinger, Heß, and Granitzer 2024). This process focuses on controlling who can access the content and under what conditions.

- *Disclosure* - Plays a focal role in reinforcing transparency and accountability. It does so by two sub-processes, Providing Content Disclaimers helps users understand that the content is generated by AI, clarifying its artificial origin and any inherent limitations (Kreps, McCain, and Brundage 2022). Simultaneously, Detecting GenAI Content involves technologies and methodologies to identify AI-generated content (Perkins et al. 2024), which is essential for ensuring the authenticity and reliability of information disseminated across digital platforms.

- *Feedback* - Occasionally overlooked, this process is integral to maintaining the quality and trustworthiness of GenAI models (Hutt et al. 2024). It involves establishing portals for user feedback and ensuring rapid responses to such feedback, which not only helps in refining the GenAI outputs but also in adhering to user expectations and regulatory standards.

### Ethics

The ethics constituent is a component that cannot be seen by itself. It was separated from the other constituents for organizational purposes of the GD. However, it is imperative to understand that the ethics of GenAI governance have to be applied into each of the previously mentioned constituents. Additionally, there are also unique processes to consider:

- *Ethical Alignment & Human Rights* - Emphasizes the alignment of GenAI technologies with human rights and intentions (Romero Moreno 2024b; Cheong, Caliskan, and Kohno 2024; Liu et al. 2022), ensuring that AI deployments do not infringe upon human dignity or freedoms. This is pivotal for fostering trust and protecting societal values.

- *Ethical Design & Deployment* - Incorporates critical sub-processes such as Ensuring Accountability & Responsibility (Bender et al. 2021) throughout the life-cycle, which highlights that GenAI should be designed with protocols to track decisions and actions for accountability. As well as, Maintaining Sustainability & Environmental Responsibility, where GenAI must be designed to minimize environmental impact, promoting sustainable practices that align with global efforts to combat environmental degradation (Wang et al. 2023b).

- *Upholding User Rights & Control* - Underscores the importance of user agency, ensuring that individuals retain control over their personal data and the decisions made by GenAI systems that affects them. This process is essential for protecting privacy and autonomy in an increasingly automated world, enabling users to have a say in how their information is used and ensuring that GenAI systems operate transparently and with user consent. However, the process is only covered by a single region. Moreover, it is important to highlight that this process could also be covered in existing data protection acts.

### Cross-Regional Comparison

Having constructed the H-GenAIGF, to assess how the different governance approaches cover the processes and sub-processes, we proceed to evaluate each region against the framework. As mentioned in the methodology, the key provisions/recommendations of each governance approach was extracted and organized under corresponding constituents. Which facilitated the comparison of the approaches. We employ symbol convention as:

- Covered (✓): A process is considered covered when all sub-processes are also covered.
- Partially Covered (-): A process is considered partially covered as long as there is one sub-process not covered or incomplete.
- Not Covered (✗): A process is considered not covered when none of the sub-processes are covered.

Our analysis reveals substantial variation in how different regions address the components outlined in our framework. This diversity reflects differing regional priorities, capacities, and strategies in tackling the multifaceted challenges posed by GenAI. For further details, we encourage readers to refer to Figure 2.

Singapore, following a mixed approach for governance of GenAI, while showing partial coverage in five out of fifteen processes and gaps in another five processes, demon-strates strong governance coverage in the realms of Ethical Alignment & Human Rights, Ethical Design & Deployment, and Model Validation & Testing. The latter, indicates a strong governance focus on ensuring that AI models are thoroughly evaluated, which is crucial for mitigating risks. Contrastingly, China's coverage for processes under the Model constituent are lacking, with only a partial coverage under Model Development, accounting for only 10% in the overall constituent. However, the coverage in the processes under the Data constituent are above 80% . Suggesting a focus on initial data handling and ethical considerations without equivalent emphasis on ongoing model governance. Moreover, it has been showed that CN's approach to govern GenAI has missing value chain aspects that may lead to accountability issues (Pi 2023).

The EU, currently ahead of requirements for GenAI governance, presents a more balanced approach with consistent, though not perfect, scores across most processes. Noteworthy is the EU's strong adherence to their risk-based approach, which allows them to cover a 92% of the processes. Similar risk-based approach is followed by US and shows a 75% of coverage amongst all processes, with six partially covered and one not covered processes. Conversely, the United Kingdom and Canada show considerable variability in their coverage, 21.8% and 31% respectively. Both approaches exhibit notable coverage gaps. On one hand, the UK covering only one, partially covering seven and not covering six. On the other hand, CA covers three, partially covers four, and does not cover eight processes. Due to the constant evolving nature of GenAI governance approaches is hard to concretely affirm that the outcome-based approach and principle-based approach are inferior. However, our analysis show that indeed presents a sizeable disadvantage in comparison to other approaches.

The stark differences in the coverage percentages underscore the global challenge of harmonizing GenAI governance. While all regions demonstrate a commitment to ethical considerations, as evidenced by generally high scores in the ethics constituent, there is a clear disparity in how they manage the operational aspects of GenAI, such as model development and data handling, specially when talking about secure storage of data. These differences not only reflect divergent regulatory priorities but also highlight the need for a more unified approach to GenAI governance. Consequently, the H-GenAIGF can aid in this regard as it provides with, what we consider, the minimum required processes and sub-processes to laid the foundation of GenAI governance.

This detailed comparison reveals that while strides are being made globally towards responsible GenAI governance, significant efforts are still required to bridge the gaps and foster a more cohesive and consolidated accepted framework that aligns with both technological advancements and societal norms.

## Governing Principles

The comparative analysis reveals that a harmonized approach is necessary as there are significant gaps across processes coverage. But how to ensure that the processes are met?, the answer may lay on the governing principles. All approaches have their set of principles, but often are not linked to specific processes. Hence, making it challenging to ensure that GenAI systems comply with provisions. Moreover, principles tend to cover different governing aspects depending on the constituent and processes they are applied to. In the following, the identified principles will be further explained:

- *Transparency and Accountability* - Widely recognized principles (Heimstädt and Dobusch 2020) , yet their implementation often lacks consistency. For instance, while the EU and US demonstrate strong adherence to these principles through rigorous regulatory disclosures, other regions like the UK and CA fall short, particularly in how transparently AI decision-making processes are documented and made accessible to the public or authorities. To enhance transparency, governments could mandate detailed reporting of AI system audits and decision-making processes, ensuring that these insights are accessible to all stakeholders, thereby promoting greater accountability throughout the life-cycle (Miguel, Naseer, and Inakoshi 2021).

- *Fairness and Integrity* - Are particularly linked to data (Oladoyinbo et al. 2024; Chen, Wu, and Wang 2023), the validation/moderation of content (Calleberg 2021; de Groot 2024), and ethical GenAI development and deployment (KIROVA et al. 2023). These principles show significant discrepancies in their practical application, particularly in mitigating biases and ensuring equitable AI outcomes. Moreover, regions need to implement additional mitigation assessments for toxicity generation as only SG emphasizes this sub-process. Toxicity can come in many forms, such as abusive language (Li et al. 2024) or content veracity (Xu, Fan, and Kankanhalli 2023). Consequently, as the end user has largely adopted GenAI through application like ChatGPT, mitigating toxicity plays an imperative aspect in the governance of GenAI.

- *Auditability and Interoperability* - Are also emerging as key challenges, especially with the increasing complexity of GenAI systems (Rane, Choudhary, and Rane 2023; Alaa et al. 2022). Enhanced audit trails and standardized protocols can facilitate better cross-system interaction and oversight, making it easier to track AI operations and verify compliance with established governance norms. Although, auditability applies to many of the processes explained in our framework, it is interesting to point out that US,SG and EU may be the only regions preparing or advising for external audits. In the case of the EU, for high-risk AI systems (that could include GenAI), audit procedures are in place (EU 2024a).

- *Sustainability and Privacy* -Are principles that call for a more integrated approach, ensuring that GenAI systems are developed with consideration for their long-term impact on the environment (Chien et al. 2023; Berthelot et al. 2024), and respecting individual privacy rights (Popowicz-Pazdej 2023).

- *Responsiveness* - Highlights the need for quick feedback response not only from providers to end-users, but also from government and providers. To be properly addressed, efficient feedback mechanisms should be in place. Consequently, allowing to adapt to feedback with ease in a way that permits models to adapt to changing conditions without compromising performance and ethical consideration.

## Case Study

The impact of ChatGPT on society is unprecedented with a variety of applications (United Arab Emirates 2023). For our case study, we employ ChatGPT-3.5&4 as it has massive reach and the challenges of such models are well documented (Fui-Hoon Nah et al. 2023). Our intention in this section is to apply our framework H-GenAIGF to evaluate the current state of ChatGPT on "compliance". We used both versions to illustrate different examples. For a more comprehensive view on all the examples tested on ChatGPT-3.5&4 and further information, refer to our website[2].

Applying the H-GenAIGF to a company's policies, could also be done. Given that the processes and the corresponding principles are already given by the framework, it is straight forward. First, the coverage of processes and sub-processes, which is contingent on whether there is access to the company's inner policies, guidelines, or frameworks. Without these documents, it will not be possible to assess the coverage of the governance processes. Second, each document should be analysed to cover every single process and sub-processes. Third, once a provision/recommendation that cover a particular process or sub-process is identified, then similar methodology to the one presented earlier should be followed to avoid biases or misplacement. However, for internal use the need for a third party (not belonging to the company) should be in place.

Next, we will introduce the ChatGPT's processes coverage following the four identified constituents:

1. *Data processes* - There are certain processes where we are unable to assess for, such as Data Acquisition, Data Preparation, and Storing & Sharing data. However, OpenAI states that it complies with GDPR and CCPA (OpenAI 2024b). Hence, GDPR compliance requires lawful and transparent data processing, limitation of data use to stated purposes, accuracy and minimization of collected data, secure storage, and demonstrable accountability by data controllers. Conversely, CCPA compliance focuses on providing consumers with rights such as disclosure of data practices, access to personal data, options to request deletion, and to opt-out of data sales, alongside ensuring non-discrimination for exercising these rights. However, due to the millions of users of ChatGPT, risk of data leakage could arise as the platform's own privacy policy (OpenAI 2023b) states, amongst others, that in certain circumstances personal information may be provided to third parties without the user being notified, unless that this is required by law. The implications of ethical considerations in ChatGPT, including handling data, are further documented in other works (Hua, Jin, and Jiang 2024).

2. *Model processes* - there has been little to none transparency in the mechanisms to ensure governance of these model. Not as a surprise, ChatGPT has only released information to address the protocols in place for Incident Reporting for security issues (OpenAI 2023a). In addition, ChatGPT leverages feedback from the research community to enhance their security and rewards found vulnerabilities, as showed in their "Bug Bounty Program" (OpenAI 2024a). As some regions such as the EU require high level of transparency with regards to general-purpose AI, and the provisions are scheduled to be applicable twelve months after entry into force ( European Parliament 2024), making providers of GenAI to transparently share information to comply with technical requirements. However, with little time for ChatGPT to disclose their practices to EU, it is unclear if efforts to be more transparent with the end-user will be taken. To cover other processes, such as Testing & Validation it could be done by using methods commonly adopted for deep neural networks, such as in (Yaghoubi and Fainekos 2019; Wang et al. 2019). Moreover, Development & Deployment details would boil down to whether the provider itself is willing (or forced) to share the information with the public.

3. *Content Generation processes* - ChatGPT covers Feedback, providing users option to leave feedback on the outputs by using the incident reporting option earlier discussed. Additionally, it actively provides instant feedback in conversations so that users can report bad responses. Moreover, partial coverage for Labeling Generated Content and Providing Content disclaimers are found. For other processes, specially for Content Validation & Moderation, it can be checked using black-box testing methods with test cases. There are several automated methods and test case creation methods, such as in (Viglianisi, Dallago, and Ceccato 2020; Darab and Chang 2014). By leveraging methods like these and adding our own test cases which involve toxic prompts (refer to our website) and their respective predicted outputs, this process could be better tested. When a user is trying to obtain sensitive information that might violate usage policies, ChatGPT covers the process of Protecting Against Unlawful Content by providing a message to the user indicating that it is unable to generate such request due to usage policies or unable to generate particular request such as copyrighted material.

4. *Ethics processes* - For ethical processes, as with many frameworks (Ashok et al. 2022; Floridi et al. 2018), there are a set of questions or criteria that would be assessed to see if that particular process is considered ethical or not. For example, for Ethical Design & Development, we could look at the points mentioned in (Aizenberg and Van Den Hoven 2020), which adopt human rights as top-level requirements, which would guide the design and development process. ChatGPT attempts to align safe and beneficial development of their models with human intention. In particular, focus on aligning with research and human intent is mentioned (OpenAI 2022) and focuses on three pillars. Training AI systems by using human feedback utilizing reinforcement learning (RL) from human feedback (Ouyang et al. 2022). Training AI systems to assist human evaluation using recursive reward modeling (RRM) (Leike et al. 2018), and to do alignment research where OpenAI states "no known indefinitely scalable solution to the alignment problem" (OpenAI 2022). The intent for alignment with human ethics is there, however, clear protocols for Ethical Design & Deployment would need to be clearly stated for accountability and responsibility throughout the life-cycle.

---
[2]https://sites.google.com/view/h-genaif/home

Processes coverage by ChatGPT demonstrates that the H-GenAIGF could be generalized to industry examples. Matching company's policies to the processes, the "minimum" required coverage can be executed to ensure a correct governance alignment. However, there are processes that are insufficiently addressed or entirely overlooked by ChatGPT, and significant concerns arise regarding model transparency, and content validation and moderation. Moreover, the lack of transparency in model processes could undermine trust and hinder the verification of compliance with ethical standards and legal frameworks, posing challenges in accountability, interoperability, and auditability, especially in jurisdictions with stringent AI regulations such as the EU.

Furthermore, the gaps in ethical design and deployment suggest potential risks of bias propagation, privacy breaches, and misalignment with human rights. Additionally, inadequate content validation and moderation mechanisms increase the risk of disseminating harmful or misleading information, which is particularly problematic given ChatGPT's extensive user base, and it would be a crucial compliance factor in regions such as CN, EU, and US. The deficiencies identified by following the H-GenAIGF affect different stakeholders, for policymakers, these processes coverage gaps demonstrates the necessity for more comprehensive GenAI regulations to ensure robust oversight and accountability. For developers, it may shows a lack of documentation on the transparency and ethical safeguards in their models which will decrease auditability on how they align with emerging regulations and societal expectations. Meanwhile, end-users face eminent risks associated with toxic generations and privacy, underscoring the urgent need for enhanced protective measures within companies to be able to fully adhere to region's governance processes.

## Limitations & Improvements

Although a comprehensive study was presented, we acknowledge that there are aspects that can be improved. For starters, new approaches are likely to come within this year, but due to the date when this work was produced, it does not cover future unseen governance processes arising from other approaches. For example, Japan's regulations for appropriate use of GenAI, which is on development. Moreover, to identify the impact of the H-GenAIGF, conducting a survey amongst AI governance connoisseurs will serve as direct feedback on the presented processes and sub-processes. In addition, to better guide GenAI governance, the governing principles may be evaluated separately and quantitatively, thereby providing auditability metrics.

To further explore the generalizability of the framework and its applicability to industry, exploring how the H-GenAIGF cover company's governance guidelines may further contribute with expanding our work. Although we studied this in a single company, as presented in our case study, there are challenges assessing how good a company covers the processes. This can be attributed to a variety reason such as the lack of internal policies to refer to and the transparency level on appropriate mechanisms to adhere to the processes. However, in future iterations, the introduction of a score metric system may be employ to further assess coverage of processes, not only by governmental approaches, but also by organizations' policies.

In addition, an important aspect for enhancement that may contribute further to the regulatory landscape, will be to identify and/or design comprehensive audits that can be applied to each of our governance processes. Hence, providing further practicability to the conformity and alignment of GenAI governance processes.

## Conclusions

This paper underscores the necessity for a harmonized governance framework for GenAI, reflecting the diverse yet critical aspects of technological, ethical, and operational governance across governance processes. The H-GenAIGF serves as a blueprint for policymakers, industry leaders, and other stakeholders to implement comprehensive and adaptive governance strategies that address both current and future challenges. The H-GenAIGF systematically articulates the following foundational components: the categorization of distinct GenAI governance processes and sub-processes, the delineation of governing principles for said processes/sub-processes, a cross-regional comparative analysis, and a case study. Collectively, these elements enhance the convoluted discourse on GenAI governance, offering a unique perspective that is both comprehensive and contextually adaptive.

Moreover, the findings of this study suggest that there are significant gaps on the coverage of the identified governance processes. While risk-based approaches have demonstrated efficacy in covering the majority of the processes, both principle-based and outcome-based approaches exhibit notable deficiencies in terms of comprehensive coverage. Conversely, hybrid approaches, such as the risk-principle approach, exhibit promising implications for a more nuanced governance structure. On the other hand, rule-based approaches, although rigorous, tend to present a more restrictive perspective that highlights a deficiency in covering crucial model processes. By fostering international cooperation and standardization, we can ensure that GenAI governance processes are covered to promote GenAI technologies that are safe, ethical, and beneficial to society at large.

## References

European Parliament. 2024. Artificial Intelligence Act: MEPs adopt landmark law. https://www.europarl.europa.eu/news/en/press-room/20240308IPR19015/artificial-

intelligence-act-meps-adopt-landmark-law. Accessed: 2024-03-13.

Abbott, R. 2020. Artificial intelligence, big data and intellectual property: protecting computer generated works in the United Kingdom. In *Research handbook on intellectual property and digital technologies*, 322–337. Edward Elgar Publishing.

Aizenberg, E.; and Van Den Hoven, J. 2020. Designing for human rights in AI. *Big Data & Society*, 7(2): 2053951720949566.

Alaa, A.; Van Breugel, B.; Saveliev, E. S.; and van der Schaar, M. 2022. How faithful is your synthetic data? sample-level metrics for evaluating and auditing generative models. In *International Conference on Machine Learning*, 290–306. PMLR.

Andreotta, A. J.; Kirkham, N.; and Rizzi, M. 2022. AI, big data, and the future of consent. *Ai & Society*, 37(4): 1715–1728.

Arya, V.; Bellamy, R. K.; Chen, P.-Y.; Dhurandhar, A.; Hind, M.; Hoffman, S. C.; Houde, S.; Liao, Q. V.; Luss, R.; Mojsilović, A.; et al. 2019. One explanation does not fit all: A toolkit and taxonomy of ai explainability techniques. *arXiv preprint arXiv:1909.03012*.

Ashok, M.; Madan, R.; Joha, A.; and Sivarajah, U. 2022. Ethical framework for Artificial Intelligence and Digital technologies. *International Journal of Information Management*, 62: 102433.

Balan, K.; Agarwal, S.; Jenni, S.; Parsons, A.; Gilbert, A.; and Collomosse, J. 2023. Ekila: synthetic media provenance and attribution for generative art. In *Proceedings of the IEEE/CVF Conference on Computer Vision and Pattern Recognition*, 913–922.

Baldassarre, M. T.; Caivano, D.; Fernandez Nieto, B.; Gigante, D.; and Ragone, A. 2023. The social impact of generative ai: An analysis on chatgpt. In *Proceedings of the 2023 ACM Conference on Information Technology for Social Good*, 363–373.

Beale, S.; Duffy, S.; Glanville, J.; Lefebvre, C.; Wright, D.; McCool, R.; Varley, D.; Boachie, C.; Fraser, C.; Harbour, J.; et al. 2014. Choosing and using methodological search filters: searchers' views. *Health Information & Libraries Journal*, 31(2): 133–147.

Bell; Burguess; Thomas; and Sadiq. 2023. Rapid Response Information Report: Generative AI - language models (LLMs) and multimodal foundation models (MFMs). Report, Australian Council of Learned Academies.

Bender, E. M.; Gebru, T.; McMillan-Major, A.; and Shmitchell, S. 2021. On the dangers of stochastic parrots: Can language models be too big? In *Proceedings of the 2021 ACM conference on fairness, accountability, and transparency*, 610–623.

Berthelot, A.; Caron, E.; Jay, M.; and Lefèvre, L. 2024. Estimating the environmental impact of Generative-AI services using an LCA-based methodology. *Procedia CIRP*, 122: 707–712.

Bird, C.; Ungless, E.; and Kasirzadeh, A. 2023. Typology of risks of generative text-to-image models. In *Proceedings of the 2023 AAAI/ACM Conference on AI, Ethics, and Society*, 396–410.

Birkstedt, T.; Minkkinen, M.; Tandon, A.; and Mäntymäki, M. 2023. AI governance: themes, knowledge gaps and future agendas. *Internet Research*, 33(7): 133–167.

Borovicka, T.; Jirina Jr, M.; Kordik, P.; and Jirina, M. 2012. Selecting representative data sets. *Advances in data mining knowledge discovery and applications*, 12: 43–70.

Calleberg, E. 2021. Making Content Moderation Less Frustrating: How Do Users Experience Explanatory Human and AI Moderation Messages.

Chen, P.; Wu, L.; and Wang, L. 2023. AI fairness in data management and analytics: A review on challenges, methodologies and applications. *Applied sciences*, 13(18): 10258.

Cheong, I.; Caliskan, A.; and Kohno, T. 2024. Safeguarding human values: rethinking US law for generative AI's societal impacts. *AI and Ethics*, 1–27.

Chien, A. A.; Lin, L.; Nguyen, H.; Rao, V.; Sharma, T.; and Wijayawardana, R. 2023. Reducing the Carbon Impact of Generative AI Inference (today and in 2035). In *Proceedings of the 2nd Workshop on Sustainable Computer Systems*, 1–7.

Choi, J. P.; Jeon, D.-S.; and Kim, B.-C. 2019. Privacy and personal data collection with information externalities. *Journal of Public Economics*, 173: 113–124.

Cihon, P. 2019. Standards for AI governance: international standards to enable global coordination in AI research & development. *Future of Humanity Institute. University of Oxford*, 340–342.

Costa, J. C.; Roxo, T.; Proença, H.; and Inácio, P. R. 2024. How deep learning sees the world: A survey on adversarial attacks & defenses. *IEEE Access*.

Cyberspace Administration of China. 2023. Measures for the Administration of Generative Artificial Intelligence Services. https://www.cac.gov.cn/2023-04/11/c_1682854275475410.htm. Accessed: Accessed: 2024-01-02.

Dafoe, A. 2018. AI governance: a research agenda. *Governance of AI Program, Future of Humanity Institute, University of Oxford: Oxford, UK*, 1442: 1443.

Darab, M. A. D.; and Chang, C. K. 2014. Black-box test data generation for gui testing. In *2014 14th International Conference on Quality Software*, 133–138. IEEE.

Davis, S. E.; Emb´ı, P. J.; and Matheny, M. E. 2024. Sustainable deployment of clinical prediction tools—a 360° approach to model maintenance. *Journal of the American Medical Informatics Association*, ocae036.

de Groot, J. 2024. Electoral integrity is at stake in Super Election Year 2024. *Atlantisch Perspectief*, 48(1): 4–8.

Dellermann, D.; Calma, A.; Lipusch, N.; Weber, T.; Weigel, S.; and Ebel, P. 2021. The future of human-AI collaboration: a taxonomy of design knowledge for hybrid intelligence systems. *arXiv preprint arXiv:2105.03354*.

Di Porto, F. 2023. Algorithmic disclosure rules. *Artificial Intelligence and Law*, 31(1): 13–51.

Dinzinger, M.; Heß, F.; and Granitzer, M. 2024. A Survey of Web Content Control for Generative AI. *arXiv preprint arXiv:2404.02309*.

Doshi, A. R.; and Hauser, O. 2023. Generative artificial intelligence enhances creativity. *Available at SSRN*.

DSIT. 2023. A pro-innovation approach to AI regulation. https://www.gov.uk/government/publications/ai-regulation-a-pro-innovation-approach. Accessed: 2023-09-01.

Du, X.; Xie, X.; Li, Y.; Ma, L.; Liu, Y.; and Zhao, J. 2019. Deepstellar: Model-based quantitative analysis of stateful deep learning systems. In *Proceedings of the 2019 27th ACM Joint Meeting on European Software Engineering Conference and Symposium on the Foundations of Software Engineering*, 477–487.

Erdélyi, O. J.; and Goldsmith, J. 2018. Regulating artificial intelligence: Proposal for a global solution. In *Proceedings of the 2018 AAAI/ACM Conference on AI, Ethics, and Society*, 95–101.

EU. 2019. Ethics guidelines for trustworthy AI. https://digital-strategy.ec.europa.eu/en/library/ethics-guidelines-trustworthy-ai. Accessed: 2024-01-12.

EU. 2024a. Artificial Intelligence Act. https://www.europarl.europa.eu/doceo/document/TA-9-2024-0138_EN.pdf. Accessed: 2024-03-14.

EU. 2024b. Living guidelines on the responsible use of generative AI in research. https://research-and-innovation.ec.europa.eu/document/2b6cf7e5-36ac-41cb-aab5-0d32050143dc_en. Accessed: 2024-03-22.

Faal, F.; Schmitt, K.; and Yu, J. Y. 2023. Reward modeling for mitigating toxicity in transformer-based language models. *Applied Intelligence*, 53(7): 8421–8435.

Ferrara, E. 2024. GenAI against humanity: Nefarious applications of generative artificial intelligence and large language models. *Journal of Computational Social Science*, 1–21.

Figoli, F. A.; Mattioli, F.; and Rampino, L. 2022. *Artificial intelligence in the design process: The Impact on Creativity and Team Collaboration*. FrancoAngeli.

Floridi, L.; Cowls, J.; Beltrametti, M.; Chatila, R.; Chazerand, P.; Dignum, V.; Luetge, C.; Madelin, R.; Pagallo, U.; Rossi, F.; et al. 2018. AI4People—an ethical framework for a good AI society: opportunities, risks, principles, and recommendations. *Minds and machines*, 28: 689–707.

Fui-Hoon Nah, F.; Zheng, R.; Cai, J.; Siau, K.; and Chen, L. 2023. Generative AI and ChatGPT: Applications, challenges, and AI-human collaboration.

G7. 2023. Hiroshima Process International Guiding Principles for Organizations Developing Advanced AI systems. https://www.mofa.go.jp/files/100573471.pdf. Accessed: 2023-11-05.

Gmyrek, P.; Berg, J.; and Bescond, D. 2023. Generative AI and jobs: A global analysis of potential effects on job quantity and quality. *ILO Working Paper*, 96.

Goodfellow, I.; Pouget-Abadie, J.; Mirza, M.; Xu, B.; Warde-Farley, D.; Ozair, S.; Courville, A.; and Bengio, Y. 2014. Generative adversarial nets. *Advances in neural information processing systems*, 27.

Gordon, G.; Rieder, B.; and Sileno, G. 2022. On mapping values in AI governance. *Computer Law & Security Review*, 46: 105712.

Government of Canada. 2022. The Artificial Intelligence and Data Act. https://ised-isde.canada.ca/site/innovation-better-canada/en/artificial-intelligence-and-data-act-aida-companion-document#s1. Accessed: 2023-11-02.

Government of Canada. 2023. Voluntary Code of Conduct on the Responsible Development and Management of Advanced Generative AI Systems. https://ised-isde.canada.ca/site/ised/en/voluntary-code-conduct-responsible-development-and-management-advanced-generative-ai-systems. Accessed: 2024-01-14.

Gupta, M.; Akiri, C.; Aryal, K.; Parker, E.; and Praharaj, L. 2023. From chatgpt to threatgpt: Impact of generative ai in cybersecurity and privacy. *IEEE Access*.

Hacker, P.; Engel, A.; and Mauer, M. 2023. Regulating ChatGPT and other large generative AI models. In *Proceedings of the 2023 ACM Conference on Fairness, Accountability, and Transparency*, 1112–1123.

Hadi, M. U.; Qureshi, R.; Shah, A.; Irfan, M.; Zafar, A.; Shaikh, M. B.; Akhtar, N.; Wu, J.; Mirjalili, S.; et al. 2023. Large language models: a comprehensive survey of its applications, challenges, limitations, and future prospects. *Authorea Preprints*.

Hamed, A. A.; Zachara-Szymanska, M.; and Wu, X. 2024. Safeguarding authenticity for mitigating the harms of generative AI: Issues, research agenda, and policies for detection, fact-checking, and ethical AI. *IScience*.

Hamid, A.; Samidi, H. R.; Finin, T.; Pappachan, P.; and Yus, R. 2023. GenAIPABench: A benchmark for generative AI-based privacy assistants. *arXiv preprint arXiv:2309.05138*.

Hankins, E.; Nettel, P. F.; Martinescu, L.; Grau, G.; and Rahim, S. 2023. Government AI Readiness Index 2023. Report, Oxford Insights.

Hannon, B.; Kumar, Y.; Sorial, P.; Li, J. J.; and Morreale, P. 2023. From Vulnerabilities to Improvements-A Deep Dive into Adversarial Testing of AI Models. In *2023 Congress in Computer Science, Computer Engineering, & Applied Computing (CSCE)*, 2645–2649. IEEE.

Hartvigsen, T.; Gabriel, S.; Palangi, H.; Sap, M.; Ray, D.; and Kamar, E. 2022. Toxigen: A large-scale machine-generated dataset for adversarial and implicit hate speech detection. *arXiv preprint arXiv:2203.09509*.

Heimstädt, M.; and Dobusch, L. 2020. Transparency and accountability: Causal, critical and constructive perspectives. *Organization Theory*, 1(4): 2631787720964216.

Hernández-Orallo, J. 2017. Evaluation in artificial intelligence: from task-oriented to ability-oriented measurement. *Artificial Intelligence Review*, 48: 397–447.

Hua, S.; Jin, S.; and Jiang, S. 2024. The Limitations and Ethical Considerations of ChatGPT. *Data Intelligence*, 6(1): 201–239.

Hutt, S.; DePiro, A.; Wang, J.; Rhodes, S.; Baker, R. S.; Hieb, G.; Sethuraman, S.; Ocumpaugh, J.; and Mills, C. 2024. Feedback on Feedback: Comparing Classic Natural Language Processing and Generative AI to Evaluate Peer Feedback. In *Proceedings of the 14th Learning Analytics and Knowledge Conference*, 55–65.

IAPP. 2024. Global AI Law and Policy Tracker. Report, International Association of Privacy Professionals.

IEEE. 2019. IEEE Draft Standard for Responsible AI Licensing. https://www.iso.org/committee/6794475.html. Accessed: 2024-01-05.

IMDA; and AIVF. 2024. PROPOSED MODEL AI GOVERNANCE FRAMEWORK FOR GENERATIVE AI. https://aiverifyfoundation.sg/downloads/Proposed_MGF_Gen_AI_2024.pdf. Accessed: 2024-01-17.

ISO. 2017. ISO/IEC JTC 1/SC 42. https://www.iso.org/committee/6794475.html. Accessed: 2024-01-15.

Jesus, V.; and Mustare, S. 2019. I did not accept that: Demonstrating consent in online collection of personal data. In *Trust, Privacy and Security in Digital Business: 16th International Conference, TrustBus 2019, Linz, Austria, August 26–29, 2019, Proceedings 16*, 33–45. Springer.

Kilovaty, I. 2025. Hacking Generative AI. *Loyola of Los Angeles Law Review*, 58.

KIROVA, V. D.; Ku, C.; Laracy, J.; and Marlowe, T. 2023. The Ethics of Artificial Intelligence in the Era of Generative AI. *Journal of Systemics, Cybernetics and Informatics*, 21(4): 42–50.

Kreps, S.; McCain, R. M.; and Brundage, M. 2022. All the news that's fit to fabricate: AI-generated text as a tool of media misinformation. *Journal of experimental political science*, 9(1): 104–117.

Lagran, E.; Searson, M.; and Trumble, J. 2024. Transforming Teacher Education in the Age of Generative AI. *Exploring New Horizons: Generative Artificial Intelligence and Teacher Education*, 2.

Laux, J.; Wachter, S.; and Mittelstadt, B. 2024. Three pathways for standardisation and ethical disclosure by default under the European Union Artificial Intelligence Act. *Computer Law & Security Review*, 53: 105957.

Leike, J.; Krueger, D.; Everitt, T.; Martic, M.; Maini, V.; and Legg, S. 2018. Scalable agent alignment via reward modeling: a research direction. *arXiv preprint arXiv:1811.07871*.

Li, L.; Fan, L.; Atreja, S.; and Hemphill, L. 2024. "HOT" ChatGPT: The promise of ChatGPT in detecting and discriminating hateful, offensive, and toxic comments on social media. *ACM Transactions on the Web*, 18(2): 1–36.

Lim, W. M.; Gunasekara, A.; Pallant, J. L.; Pallant, J. I.; and Pechenkina, E. 2023. Generative AI and the future of education: Ragnarök or reformation? A paradoxical perspective from management educators. *The international journal of management education*, 21(2): 100790.

Liu, R.; Zhang, G.; Feng, X.; and Vosoughi, S. 2022. Aligning generative language models with human values. In *Findings of the Association for Computational Linguistics: NAACL 2022*, 241–252.

McGregor, S. 2021. Preventing repeated real world AI failures by cataloging incidents: The AI incident database. In *Proceedings of the AAAI Conference on Artificial Intelligence*, volume 35, 15458–15463.

Miguel, B. S.; Naseer, A.; and Inakoshi, H. 2021. Putting accountability of AI systems into practice. In *Proceedings of the Twenty-Ninth International Conference on International Joint Conferences on Artificial Intelligence*, 5276–5278.

Mökander, J.; Axente, M.; Casolari, F.; and Floridi, L. 2022. Conformity assessments and post-market monitoring: a guide to the role of auditing in the proposed European AI regulation. *Minds and Machines*, 32(2): 241–268.

Myers, D.; Mohawesh, R.; Chellaboina, V. I.; Sathvik, A. L.; Venkatesh, P.; Ho, Y.-H.; Henshaw, H.; Alhawawreh, M.; Berdik, D.; and Jararweh, Y. 2024. Foundation and large language models: fundamentals, challenges, opportunities, and social impacts. *Cluster Computing*, 27(1): 1–26.

NIST. 2024. Artificial Intelligence Risk Management Framework: Generative Artificial Intelligence Profile. https://www.nist.gov/itl/ai-risk-management-framework. Accessed: 2024-05-02.

OECD. 2024. AI Principles. https://oecd.ai/en/ai-principles. Accessed: 2024-05-04.

Oladoyinbo, T. O.; Olabanji, S. O.; Olaniyi, O. O.; Adebiyi, O. O.; Okunleye, O. J.; and Ismaila Alao, A. 2024. Exploring the challenges of artificial intelligence in data integrity and its influence on social dynamics. *Asian Journal of Advanced Research and Reports*, 18(2): 1–23.

OpenAI. 2022. Our approach to alignment research. https://openai.com/index/our-approach-to-alignment-research/. Accessed: 2024-05-02.

OpenAI. 2023a. Coordinated vulnerability disclosure policy. https://openai.com/policies/coordinated-vulnerability-disclosure-policy/. Accessed: 2024-05-02.

OpenAI. 2023b. Privacy policy. https://openai.com/policies/privacy-policy/. Accessed: 2024-05-02.

OpenAI. 2024a. OpenAI Bug Bounty Program. https://bugcrowd.com/openai. Accessed: 2024-05-02.

OpenAI. 2024b. Security and privacy. https://openai.com/security/. Accessed: 2024-05-02.

Ouyang, L.; Wu, J.; Jiang, X.; Almeida, D.; Wainwright, C.; Mishkin, P.; Zhang, C.; Agarwal, S.; Slama, K.; Ray, A.; et al. 2022. Training language models to follow instructions with human feedback. *Advances in neural information processing systems*, 35: 27730–27744.

Pagallo, U.; Aurucci, P.; Casanovas, P.; Chatila, R.; Chazerand, P.; Dignum, V.; Luetge, C.; Madelin, R.; Schafer, B.; and Valcke, P. 2019. On good AI governance: 14 priority actions, a SMART model of governance, and a regulatory toolbox. *SSRN*, 1–49.

Partadiredja, R. A.; Serrano, C. E.; and Ljubenkov, D. 2020. AI or human: the socio-ethical implications of AI-generated media content. In *2020 13th CMI Conference on Cybersecurity and Privacy (CMI)-Digital Transformation-Potentials and Challenges (51275)*, 1–6. IEEE.

Perkins, M.; Roe, J.; Postma, D.; McGaughran, J.; and Hickerson, D. 2024. Detection of GPT-4 generated text in higher education: Combining academic judgement and software to identify generative AI tool misuse. *Journal of Academic Ethics*, 22(1): 89–113.

Pi, Y. 2023. Missing Value Chain in Generative AI Governance: China as an example. In *NeurIPS 2023 Workshop on Regulatable ML*.

Popowicz-Pazdej, A. 2023. The proportionality between trade secret and privacy protection: How to strike the right balance when designing generative AI tools. *Journal of Data Protection & Privacy*, 6(2): 153–166.

Raji, I. D.; Bender, E. M.; Paullada, A.; Denton, E.; and Hanna, A. 2021. AI and the everything in the whole wide world benchmark. *arXiv preprint arXiv:2111.15366*.

Rane, N.; Choudhary, S.; and Rane, J. 2023. Integrating Building Information Modelling (BIM) with ChatGPT, Bard, and similar generative artificial intelligence in the architecture, engineering, and construction industry: applications, a novel framework, challenges, and future scope. *Bard, and similar generative artificial intelligence in the architecture, engineering, and construction industry: applications, a novel framework, challenges, and future scope (November 22, 2023)*, 1–22.

Romero Moreno, F. 2024a. Generative AI and deepfakes: a human rights approach to tackling harmful content. *International Review of Law, Computers & Technology*, 1–30.

Romero Moreno, F. 2024b. Generative AI and deepfakes: a human rights approach to tackling harmful content. *International Review of Law, Computers & Technology*, 1–30.

Samoili, S.; Cobo, M. L.; Gómez, E.; De Prato, G.; Martínez-Plumed, F.; and Delipetrev, B. 2020. AI Watch Defining Artificial Intelligence 2.0. *na*, 1–125.

Samuelson, P. 2023. Legal challenges to generative AI, Part I. *Communications of the ACM*, 66(7): 20–23.

Schramowski, P.; Brack, M.; Deiseroth, B.; and Kersting, K. 2023. Safe latent diffusion: Mitigating inappropriate degeneration in diffusion models. In *Proceedings of the IEEE/CVF Conference on Computer Vision and Pattern Recognition*, 22522–22531.

Selva, J.; Johansen, A. S.; Escalera, S.; Nasrollahi, K.; Moeslund, T. B.; and Clapés, A. 2023. Video transformers: A survey. *IEEE Transactions on Pattern Analysis and Machine Intelligence*.

Smart Nation. 2023. Singapore National AI Strategy 2.0. https://aiverifyfoundation.sg/downloads/Proposed_MGF_Gen_AI_2024.pdf. Accessed: 2023-12-20.

Sohn, J. H.; Chillakuru, Y. R.; Lee, S.; Lee, A. Y.; Kelil, T.; Hess, C. P.; Seo, Y.; Vu, T.; and Joe, B. N. 2020. An open-source, vender agnostic hardware and software pipeline for integration of artificial intelligence in radiology workflow. *Journal of digital imaging*, 33: 1041–1046.

Sun, B.; Sun, J.; Koh, W.; and Shi, J. 2024. Neural Network Semantic Backdoor Detection and Mitigation: A Causality-Based Approach. In *Proceedings of the 33rd USENIX Security Symposium. USENIX Association, San Francisco, CA, USA*.

The White House. 2023. Executive Order on the Safe, Secure, and Trustworthy Development and Use of Artificial Intelligence. https://www.whitehouse.gov/briefing-room/presidential-actions/2023/10/30/executive-order-on-the-safe-secure-and-trustworthy-development-and-use-of-artificial-intelligence/. Accessed: 2023-11-03.

UKGov. 2024. Generative AI framework for HM Government. https://www.gov.uk/government/publications/generative-ai-framework-for-hmg. Accessed: 2024-02-15.

UNESCO. 2023a. Guidance for generative AI in education and research. https://www.unesco.org/en/articles/guidance-generative-ai-education-and-research. Accessed: 2024-01-11.

UNESCO. 2023b. Recommendation on the Ethics of Artificial Intelligence. https://example.com. Accessed: 2023-08-10.

United Arab Emirates. 2023. 100 Practical Applications and Use Cases of Generative AI. https://ai.gov.ae/wp-content/uploads/2023/04/406.-Generative-AI-Guide_ver1-EN.pdf. Accessed: 2024-02-05.

US Congress. 2023a. Candidate Voice Fraud Prohibition Act. https://www.congress.gov/bill/118th-congress/house-bill/4611/text?s=7&r=1&q=%7B%22search%22%3A%22candidate+voice+fraud+prohibition+act%22%7D. Accessed: 2024-01-20.

US Congress. 2023b. Preventing Deep Fake Scams Act. https://www.congress.gov/bill/118th-congress/house-bill/5808/text?s=6&r=1&q=%7B%22search%22%3A%22Preventing+Deep+Fake+Scams+Act%22%7D. Accessed: 2024-01-20.

US Congress. 2024a. Generative AI Copyright Disclosure Act of 2024. https://www.congress.gov/bill/118th-congress/house-bill/7913/text?s=2&r=1&q=%7B%22search%22%3A%22H.+R.+7913%22%7D. Accessed: 2024-04-20.

US Congress. 2024b. No AI FRAUD Act. https://www.congress.gov/bill/118th-congress/house-bill/6943/text?s=8&r=1&q=%7B%22search%22%3A%22No+AI+fraud+Act%22%7D. Accessed: 2024-01-20.

Vaswani, A.; Shazeer, N.; Parmar, N.; Uszkoreit, J.; Jones, L.; Gomez, A. N.; Kaiser, Ł.; and Polosukhin, I. 2017. Attention is all you need. *Advances in neural information processing systems*, 30.

Viglianisi, E.; Dallago, M.; and Ceccato, M. 2020. Resttestgen: automated black-box testing of restful apis. In *2020 IEEE 13th International Conference on Software Testing, Validation and Verification (ICST)*, 142–152. IEEE.

Wach, K.; Duong, C. D.; Ejdys, J.; Kazlauskaitė, R.; Korzynski, P.; Mazurek, G.; Paliszkiewicz, J.; and Ziemba, E. 2023. The dark side of generative artificial intelligence: A critical analysis of controversies and risks of ChatGPT. *Entrepreneurial Business and Economics Review*, 11(2): 7–30.

Wang, H.; Wu, C.; and Zheng, K. 2024. Defense against adversarial attacks based on color space transformation. *Neural Networks*, 106176.

Wang, J.; Dong, G.; Sun, J.; Wang, X.; and Zhang, P. 2019. Adversarial sample detection for deep neural network through model mutation testing. In *2019 IEEE/ACM 41st International Conference on Software Engineering (ICSE)*, 1245–1256. IEEE.

Wang, L.; Xie, X.; Du, X.; Tian, M.; Guo, Q.; Yang, Z.; and Shen, C. 2023a. DistXplore: Distribution-guided testing for evaluating and enhancing deep learning systems. In *Proceedings of the 31st ACM Joint European Software Engineering Conference and Symposium on the Foundations of Software Engineering*, 68–80.

Wang, R.; Li, C.; Li, X.; Deng, R.; and Dong, Z. 2023b. GenAI4Sustainability: GPT and Its Potentials For Achieving UN's Sustainable Development Goals. *IEEE/CAA Journal of Automatica Sinica*, 10(12): 2179–2182.

Wang, Z.; She, Q.; and Ward, T. E. 2021. Generative adversarial networks in computer vision: A survey and taxonomy. *ACM Computing Surveys (CSUR)*, 54(2): 1–38.

WEF. 2024. AI Governance Alliance Briefing Paper Series. https://www.weforum.org/publications/ai-governance-alliance-briefing-paper-series/. Accessed: 2024-01-20.

Wittenberg, C.; Epstein, Z.; Berinsky, A. J.; and Rand, D. G. 2024. Labeling AI-Generated Content: Promises, Perils, and Future Directions. https://mit-genai.pubpub.org/pub/hu71se89/release/1. Accessed: 2024-05-01.

Wu, T.; He, S.; Liu, J.; Sun, S.; Liu, K.; Han, Q.-L.; and Tang, Y. 2023. A brief overview of ChatGPT: The history, status quo and potential future development. *IEEE/CAA Journal of Automatica Sinica*, 10(5): 1122–1136.

Xie, X.; Guo, W.; Ma, L.; Le, W.; Wang, J.; Zhou, L.; Liu, Y.; and Xing, X. 2021. RNNrepair: Automatic RNN repair via model-based analysis. In *International Conference on Machine Learning*, 11383–11392. PMLR.

Xie, X.; Li, T.; Wang, J.; Ma, L.; Guo, Q.; Juefei-Xu, F.; and Liu, Y. 2022. Npc: N euron p ath c overage via characterizing decision logic of deep neural networks. *ACM Transactions on Software Engineering and Methodology (TOSEM)*, 31(3): 1–27.

Xie, X.; Ma, L.; Juefei-Xu, F.; Xue, M.; Chen, H.; Liu, Y.; Zhao, J.; Li, B.; Yin, J.; and See, S. 2019a. Deephunter: a coverage-guided fuzz testing framework for deep neural networks. In *Proceedings of the 28th ACM SIGSOFT international symposium on software testing and analysis*, 146–157.

Xie, X.; Ma, L.; Wang, H.; Li, Y.; Liu, Y.; and Li, X. 2019b. Diffchaser: Detecting disagreements for deep neural networks. In *Proceedings of the Twenty-Eighth International Joint Conference on Artificial Intelligence*, 5772–5778.

Xu, D.; Fan, S.; and Kankanhalli, M. 2023. Combating misinformation in the era of generative AI models. In *Proceedings of the 31st ACM International Conference on Multimedia*, 9291–9298.

Yaghoubi, S.; and Fainekos, G. 2019. Gray-box adversarial testing for control systems with machine learning components. In *Proceedings of the 22nd ACM International Conference on Hybrid Systems: Computation and Control*, 179–184.

Zarifhonarvar, A. 2023. Economics of chatgpt: A labor market view on the occupational impact of artificial intelligence. *Journal of Electronic Business & Digital Economics*.